\documentclass[showpacs,preprint,aps,amsmath,15pt,firstpage,superscriptaddress]{revtex4}
\usepackage[dvips]{graphicx}
\usepackage[dvips]{color}
\usepackage{subfig}

\begin{document}

\title{Enhancement of coherent energy transfer by disorder and temperature in light harvesting processes}
\author{Shi-Jie Xiong}
\email{sjxiong@nju.edu.cn}
\affiliation{School of Materials Science and Engineering, Nanyang Technological University, Singapore 639798, Singapore}
\affiliation{%
National Laboratory of Solid State Microstructures and Department
of Physics, Nanjing University, Nanjing, China
}%
\author{Ye Xiong}
\affiliation{Department of Physics, Nanjing Normal University, Nanjing,
China}
\author{Yang Zhao}
\email{YZhao@ntu.edu.sg}
\affiliation{School of Materials Science and Engineering, Nanyang Technological University, Singapore 639798, Singapore}

\date{\today}


\begin{abstract}

We investigate the influence of static disorder and thermal excitations on excitonic energy transport in the light-harvesting apparatus of photosynthetic systems by solving the Schr\"{o}dinger equation and taking into account the coherent hoppings of excitons, the rates of exciton creation and annihilation in antennas and reaction centers, and the coupling to thermally excited phonons. The antennas and reaction centers are modeled, respectively, as the sources and drains which  provide the channels for creation and annihilation of excitons. Phonon modes below a maximum frequency are coupled to the excitons that are continuously created in the antennas and depleted in the reaction centers, and the phonon population in these modes obeys the Bose-Einstein distribution at a given temperature. It is found that the energy transport is not only robust against the static disorder and the thermal noise, but it can also be enhanced by increasing the randomness and temperature in most parameter regimes. Relevance of our work to the highly efficient energy transport in photosynthetic systems is discussed.
\end{abstract}

\maketitle

\section{Introduction}

Photosynthesis is an essential means to obtain energy for all life forms on
Earth. Although the detailed structures of photosynthetic systems
are complicated and species-dependent, it is now believed that solar photons are
absorbed to produce electronic excited states (called excitons) in
molecular chromophores that are found in light-harvesting (or
antenna) complexes. Many antenna complexes, together with reaction
center complexes which can convert the energy stored in the
excitons to chemical energy via biochemical reactions,
form a transportation network for excitons. The photoexcitation process starts with the exciton creation in the antenna complexes, which is followed by exciton transfer across the pigment network, and ends with exciton trapping by the reaction center complexes. This energy transport process is completed on a 10 - 100 picosecond timescale \cite{time1,time2,time3,time4,time5,time6}. In recent years the
high efficiency and quantum coherence of the exciton transfer processes amidst a noisy environment have atrracted great interest \cite{Zhao1,Zhao2,Ygc,wn,EET,Moix,vilnus,Chin,ulrich,cohe,exp1,exp2,exp3,exp4}. The mechanism for such coherent transport has been extensively
studied with various theoretical models \cite{the1,the2,the3,the4,the5,Sj}. The exciton transfer from antennas to reaction centers is often modeled by the semiclassical F\"{o}rster theory which considers incoherent hoppings between sites \cite{the1,the2}. To account for coherence in the energy transport processes, a microscopic description is provided by the Redfield theory where the exciton dynamics is solved from a master equation in a reduced space of excitons in the weak phonon coupling and Born-Markov approximation \cite{the3}. Based on this theory, Silbey {\it et al.} developed an approach for the diffusion of excitons \cite{the4,the5}; Haken and Strobl used a stochastic model \cite{the6}; and Kenkre and Knox established a generalized master equation formalism \cite{the7,the8} to investigate the coherent and incoherent aspects of the excitonic transfer. To make the models more realistic, Zhang {\it et al.}~introduced a modified Redfield equation to include static disorder in the exciton system \cite{the9} which was later used to simulate the energy transfer dynamics in light-harvesting complexes of green plants \cite{the10}. Silbey and co-workers proposed a generalized theory which includes coherent transport effects only within donors and acceptors while treating interactions between them with the standard F\"{o}rster model \cite{the11,the12,the13}. A comparison between F\"{o}rster, Redfield, and other models is given by Yang and Fleming in Ref.~\cite{the14}. With a master equation approach in the Born-Markov and secular approximations, Mohseni {\it et al.} carried out computer simulations and found that the energy transport efficiency can be enhanced by modulating environmental noise \cite{the15}.

In the energy transfer process in photosynthetic systems, excitons are successionally created in the antennas and then trapped in the reaction centers, generating an uninterrupted energy flow from the antennas to the reaction centers.
This energy flow is significantly affected not only by the coherence and decoherence effects during the exciton transfers
across the pigment network, but also by the creation and annihilation of the excitons in the antennas and the reaction centers, respectively.
In fact, if the excitons are viewed as
energy carriers in the photosynthetic processes, the creation and
annihilation of them in the antennas and the reaction centers play roles of
source and drain of carriers, respectively. Thus, the widely used
source-network-drain models describing continuous current of
electron transport in quantum systems may also be applicable to treat the energy flow
in photosynthetic systems if
the creation and annihilation of excitons can be properly addressed.

This work is aimed to describe the excitonic energy transport in
photosynthetic systems with a full quantum-mechanical
source-network-drain model. We include on an equal footing effects of scattering,
disorder, exciton-phonon interactions, thermal excitations in a noisy environment, and efficiencies of antennas and reaction centers.
The key issues
are how to model exciton creations and annihilations with a suitable pair of
source and drain, and how to include the thermal
(temperature) effect in such a quantum-mechanical treatment. We use incoming (source) channels and outgoing (drain) channels for the creation and annihilation of excitons in antennas and reaction centers, and those channels are semi-infinite with the ends connected to different sites of a  pigment network so that the exciton current from the source to the drain can be sustained. Considering that this process is taking place in a thermal bath, we include interactions between excitons and bath phonons in the network.
With these key ingredients captured by a Hermitian Hamiltonian, the wave functions of exciton and phonons can be directly solved via the Schr\"{o}dinger equation without invoking any classical or semiclassical approximations. Structural fluctuations in realistic systems are modeled with adding disorders to control parameters in the Hamiltonian. Exciton creation and annihilation rates can be adjusted by parameter-tuning of incoming and outgoing channels. After obtaining the wave functions, the exciton current, which reflects the efficiency of the photosynthesis, is calculated via the  extended Landauer-Buttiker formula which works in the Hilbert space for many-body states \cite{lb,xx1,xx2}. The numerical results show that the creation and annihilation rates in antennas and reaction centers and their adaptability have crucial effects on the total efficiency and the degree of coherence in the network. It is interesting that the efficiency may be enhanced by introducing randomness and thermal excitations in many parameter regimes. At ambient temperature the efficiency may reach its maximum or saturation. This can help explain why photosynthesis is so efficient at room temperature and in noisy environments.

The paper is organized as follows. In the next section we present the basic model and formalism. In the third section we discuss the results in the absence of interactions with phonons and analyze the effect of static randomness. In section IV we include the interaction with phonons and investigate the effect of thermal excitations on the energy transport. A brief summary is given in section V.

\section{Model and formalism}

We consider the following Hamiltonian for the exciton transport in a photosynthetic energy-transfer system:
\begin{equation}
    H_0 =  H_c + H_s +H_d,
    \end{equation}
where $H_c$ is the Hamiltonian of a pigment network for exciton transport, such as the the light harvesting LH1
and LH2 photosynthetic complexes in a membrane of {\it Rsp. photometricum} \cite{lh1,lh2} and the Fenna-Mathews-Olsen (FMO) protein complex in green sulfur bacteria \cite{fmo}, connecting the antennas with the reaction centers.
$H_c$ can be written as
\begin{equation}
   H_c = \sum_{i} \epsilon_i a^\dag_{i,0} a_{i,0} + \sum_{i \neq j}
   J _{i,j} a^\dag_{i,0} a_{j,0},
   \end{equation}
where $a^\dag_{i,0}$ is the creation operator of exciton on site $i$,
$\epsilon_i$ is the site energy for excitons, and $J_{i,j}$ is the hopping integral
for excitons from site $j$ to site $i$. One has $J_{i,j} =
J^*_{j,i}$ as the Hamiltonian is Hermitian. Here a site may label a single chromophore or a cluster of chromophores. For the former,  $J_{i,j}$ originates from nearest-neighbor transfer integrals and F\"{o}rster dipole-dipole interactions between well-separated pigments, both of which are strongly dependent on the distance between the pigments \cite{the1,fd2,Zhao_PRE,ulrich,Busch}. For the latter, $\epsilon_i$  is the resonance level to host an exciton in the $i$th cluster of chromophores, while $J_{i,j}$ labels the effective transfer integral between $i$th and $j$th clusters with a distance dependence that differs from that in the F\"{o}rster theory \cite{the11}.
$H_s$ ($H_d$) is the tight-binding
Hamiltonian for the source (drain), describing a semi-infinite
virtual chain connected to a site in the pigment network as a
path of the successional creation (annihilation) of excitons in an
antenna (reaction center) by radiation (biochemical reaction),
\begin{equation}
  H_{s(d)} = \sum_{i \in {\cal S}({\cal D})}
  \sum_{n=1}^{\infty} [ g_i (a^\dag_{i,n} a_{i,n-1} +a^\dag_{i,n-1}a_{i,n}) +
  \epsilon_i a^\dag_{i,n}
  a_{i,n} ],
  \end{equation}
where a semi-infinite chain is labeled by the index $i$ of a network site to which it is connected, ${\cal S}$ (${\cal D}$) is the set of chains belonging to source (drain), $n$ denotes the position on the chain counting from the pigment network, and $g_i$ is the
nearest-neighbor hopping integral in the $i$th virtual chain.
Thus the tight-binding Hamiltonian of the $i$th chain yields a transition exciton band of width $4g_i$ centered at $\epsilon_i$, and both $g_i$ and $\epsilon_i$ vary from chain to chain as the transition bands depend on detailed configurations. The widths of transition bands in the virtual chains mimic the energy uncertainty of excitons caused by the timescales of the photon-exciton conversion events in the antennas and charge separation events in the reaction centers. For systems relevant to photosynthesis \cite{castro,ww1}, $g_i$ varies from 1 meV to 10 meV, and $\epsilon_i$ from 1 eV to 2 eV. As the events occur in a quantized manner, these band widths are intrinsic parameters independent of actual light intensity in antennas or actual energy flow in reaction canters. In fact, the widths only set upper limits on the exciton currents flowing from the sources to the drains, and actual currents may be much smaller than those limits as they are also determined by light intensities and adsorption cross sections in antennas and by exciton trapping rates in reaction centers \cite{lh1}. These additional physical ingredients, which can be viewed as external conditions that vary from time to time, are not included in the Hamiltonian, but their influences on exciton statistics in the sources and the drains will be taken into account in the formulation of the exciton current.
The combined Hamiltonian may be illustrated by Fig.~1, where the networked sites corresponding to the antennas
(reaction centers) are represented with green (magenta) circles, and the creation
(annihilation) of excitons is represented with dashed incoming
(dotted outgoing) arrows attached to corresponding sites.
For a specific system such as FMO, a site in the pigment network may be connected to an antenna or a reaction center, and it is also possible that it is linked only to adjacent sites in the network. Detailed configurations in the source-network-drain model can be determined from a realistic photosynthetic system of interest. In this work, however, we are concerned with generic features of energy transfer in photosynthesis instead of effects of detailed configurations. We will therefore consider a network in two dimensions (2D) with closely packed sites with a distribution of control parameters. Static disorder erases structural details but the salient features remain after configurational averages. The 2D network to be studied here resembles, for example, the RC-LH1-LH2 photosynthetic complexes in a membrane of {\it Rsp. photometricum}.

\begin{figure}
\includegraphics[width=13.5cm]{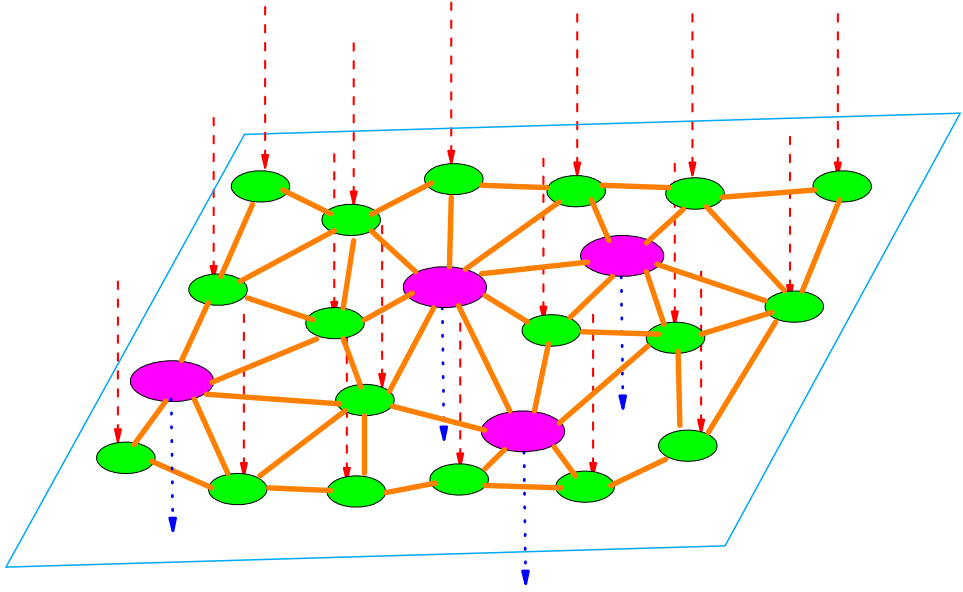}
\caption[FIG]{\label{FIG.1} Schematic for a network of exciton transport. The
antenna (reaction center) sites are represented with green (magenta)
circles which are interconnected forming a network, and
creation (annihilation) of excitons is represented with dashed
incoming (dotted outgoing) arrows attached to corresponding sites.
The couplings between sites, represented with brown links in the network, correspond to hopping integrals $J_{i,j}$ in Hamiltonian $H_c$, while the incoming and outgoing channels expressed by the arrows depict virtual semi-infinite chains in Hamiltonians $H_s$ and $H_d$. }
\end{figure}

In the photosynthetic process the excitons are constantly flowing into the network via the incoming channels.
The incident exciton with energy $E$ in an incoming channel $i$ ($i \in {\cal S}$) can be expressed by a plane wave propagating along channel $i$ towards the network
\begin{equation}
  \psi_{i}^{\rm in}(E) = \sum_{n=0}^\infty \exp[-i k_i(E) n] a^\dag_{i,n} | 0\rangle,
  \end{equation}
where $k_i (E) = \arccos [({E-\epsilon_i})/{2 g_i} ]$ is the wave vector and $E$ is distributed within the interval $[\epsilon_i -2g_i, \epsilon_i+2g_i]$. The coefficients in the summation are set to unity implying that the rate of incident flow is ${4g_i}/{h}$ if the transition band in the $i$th chain is occupied by excitons. The incoming wave propagates around in the network and then either exits via the outgoing channels or gets reflected to the incoming channels. The outgoing part and reflected part of the wave function are also expressed with plane waves in corresponding channels:
\begin{equation}
  \psi_{ij }^{\rm out}(E) = \sum_{n=0}^\infty t_{ij}(E) \exp[i k_j(E) n] a^\dag_{j,n} | 0\rangle \,\,\,{\rm   for   }\,\,\, j \in {\cal D},
  \end{equation}
and
\begin{equation}
  \psi_{i,i' }^{\rm ref}(E) = \sum_{n=0}^\infty r_{ii'}(E)\exp[i k_{i'}(E) n] a^\dag_{i',n} | 0\rangle \,\,\, {\rm  for  } \,\,\,  i' \in {\cal S},
  \end{equation}
where $t_{ij}$ and $r_{ii'}$ are transmission and reflection amplitudes from incident channel $i$ to channel $j$ ($ \in {\cal D}$) and to channel $i'$ ($\in {\cal S}$), respectively. The complete form of the wave function for exciton with energy $E$ and entering via channel $i$ is
\begin{equation}
\label{wave}
  \psi_i (E) =  \psi_{i}^{\rm in}(E) + \sum_{j \in {\cal D}} \psi_{ij}^{\rm out}(E)
  +\sum_{i' \in {\cal S}} \psi_{ii'}^{\rm ref} (E).
  \end{equation}
Note that the wave function is normalized according to the incident rate.

The transmission and reflection amplitudes $t_{ij}(E)$ and $r_{ii'}(E)$ can be solved from the Sch\"{o}dinger equation with the transfer-matrix or Green function
technique,
\begin{equation}
     H_0 \psi_i (E) = E \psi_i (E).
\end{equation}
The total exciton current through the system, representing the
total efficiency of the photosynthesis, can be calculated by the Landauer-Buttiker formula summing over all incident and output channels and integrating over energy \cite{lb} as:
\begin{equation}
   I= \frac{1}{h} \sum_{ i\in {\cal S}, j \in
    {\cal D}} \int_{{\cal E}_{ij}} dE \left[ \frac{ |t_{ij}(E) |^2 \eta_{ij}(E) v_j(E)}{v_i(E)}
    -\frac{ |t_{ji}(E) |^2 \eta_{ji}(E) v_i(E)}{v_j(E)} \right].
    \end{equation}
Here $|t_{ij}(E)|^2$ is the transmission probability from channel $i$ to
channel $j$ for given energy $E$,
$v_i(E)$ is the exciton velocity in the $i$th chain, and the ratio ${ v_j(E)}/v_i(E)$ accounts for the difference in transport rate between the input and output channels.
The integration range ${\cal E}_{ij}$ is the region of $E$ where both
$k_i(E)$ and $k_j(E)$ are real. The first and second terms in the integrand correspond to the currents from source to drain (positive) and from drain to source (negative), respectively. The sign of the total current is controlled by $\eta_{ij}(E)$, the probability that the state of energy $E$ in the $i$th chain has exciton to emit while the state in the $j$th chain is empty and available to adsorb the exciton. So we can express $\eta_{ij}(E)$ as $ \eta_{ij}(E) = p_i(E)q_j(E)$, where $p_i(E)$ ($q_j(E)$) is probability of state in the $i$th ($j$th) chain being occupied (vacant). As the excitons are created (adsorbed) only in source (drain) chains, $p_i(E)$ ($q_j(E)$) is nonzero only when $i \in {\cal S}$ ($j \in {\cal D}$). This guarantees that the second term in the above formula is always zero and the current is positive (from source to drain). $p_i(E)$ as a function of energy is determined by the light intensity and adsorption cross section in antennas, while $q_j(E)$ depends on the reopening timescale in reaction centers \cite{lh1}. Although $p_i(E)$ and $q_j(E)$ set a further limitation on the energy transport, their effect can be taken into account simply by introducing a prefactor to the current. For the sake of simplicity, we will adopt a simple form of $\eta_{ij}(E)$:
\begin{equation}
\label{eta}
  \eta_{ij}(E) =\left\{ \begin{array}{l} 1, \text{   for   } i \in {\cal S} \text{   and   } j \in {\cal D},\\ 0, \text{   otherwise.}  \end{array} \right.
  \end{equation}
Then the current can be calculated as
\begin{equation}
    I= \frac{1}{h} \sum_{ i\in {\cal S}, j \in
    {\cal D}} \int_{{\cal E}_{ij}} dE \frac{ |t_{ij}(E) |^2 |g_j \sin
    k_j(E)|}{|g_i \sin k_i(E)|},
    \end{equation}
    where $|g_j \sin k_j(E)|/|g_i \sin k_i(E)|$ is the ratio of velocities in channels $i$ and $j$.

Taking into account the configurational fluctuations in the network, the antennas, and the reaction centers, we introduce disorder in $\epsilon_i$ and $J_{ij}$. Since the inter-site coupling in the network decreases rapidly as the inter-site distance increases, only couplings between nearest neighbors (NN) are kept. The parameter distributions can then be written as
\begin{equation}
  P(\epsilon_i) = \left\{ \begin{array}{l} 1/w_{s(d)}, \,\, \text{for} \,\,
  i \in {\cal S} ({\cal D}) \,\, \text{ and } \, \epsilon_{s(d)}-w_{s(d)}/2 \leq
  \epsilon_i \leq \epsilon_{s(d)}+w_{s(d)}/2, \\ 0, \text{  otherwise,}
  \end{array} \right.
  \end{equation}
\begin{equation}
  P(J_{ij}) = \left\{ \begin{array}{l} 1/W, \,\, \text{for} \,\,
  i,j  \text{ are NN sites } \,\, \text{ and } \, J_0-W/2 \leq
  J_{ij} \leq J_0+W/2, \\ 0, \text{  otherwise.}
  \end{array} \right.
  \end{equation}
Here $w_{s(d)} $ and $W$ quantify disorder in $\epsilon_i$ for $i \in {\cal S} ({\cal D}) $ and in $J_{ij}$ for NN sites, respectively, while $\epsilon_{s(d)}$ and $J_0$ are their averages. On the other hand, in order to reduce the number of parameters, we set $g_i = g_s$ for $i \in {\cal S}$ and $g_i =g_d$ for $i \in {\cal D}$.

\section{Effect of static disorder in the absence of interaction with phonons}

In this section we investigate general features of energy transfers from the source to the drain by taking into account static disorder in the pigment network which reflects configurational randomness of the photosynthetic system. We carry out numerical calculations for a network of
400 sites closely packed in a 2D plane. At this stage we do not include interactions with phonons and other thermal excitations, and focus instead on how the efficiency of energy transfer is affected by configurations and disorder of the network. Also randomly distributed in this network are the source sites connected to the antennas and the drain sites connected to the reaction centers. The density of the drain sites is denoted
by $p$. In order to reduce the number of parameters in our model, it is assumed that each site in the network is connected either to an antenna or a reaction center, so the probabilities of a site being source and drain are $1-p$ and $p$, respectively. By simply removing some incoming or outgoing channels, our calculation can be easily extended to include cases where some sites in the pigment network are connected to neither antennas nor reaction centers.

\begin{figure}
\includegraphics[width=13.5cm]{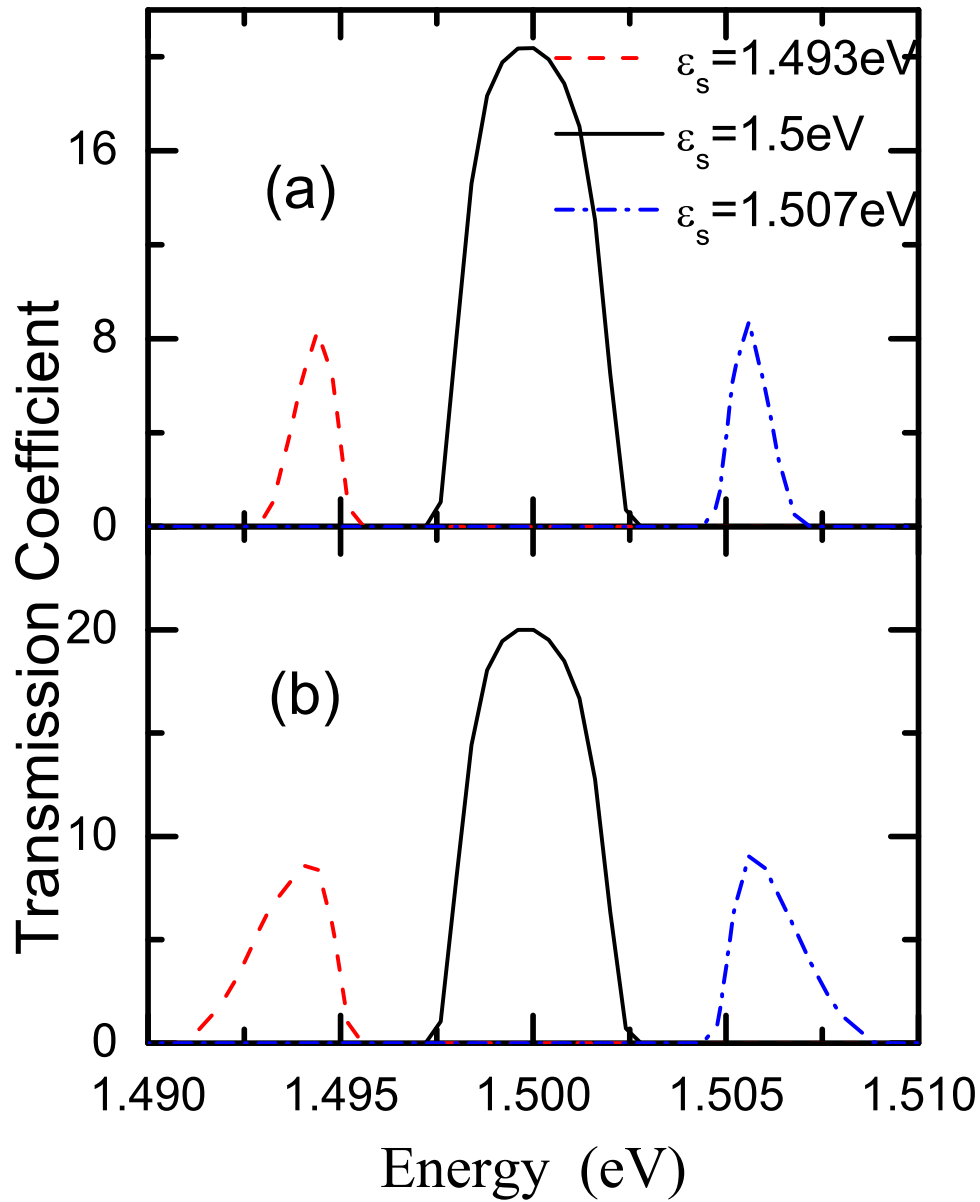}
\caption[FIG]{\label{FIG.2} Total transmission coefficient as a
function of the photon energy. (a) $w_d=0.002$eV, (b) $w_d=0.006$eV. Other parameters
are: $\epsilon_d =1.5$eV, $w_s =0.001$eV, $g_d=0.003$eV, $g_s=0.001$eV, $J_0=0.01$eV, $W=0.005$eV,
and $p =0.3$. For a given set of parameters, there is only a single peak corresponding to the global resonance of the entire source-network-drain system.}
\end{figure}

Figure~2 displays the network transmission spectrum, i.e., the total
probability of energy transfer from the incoming photons to the reaction centers as a function of the photon energy,
for two different amplitudes of drain disorder.
The spectrum has a global resonance peak of the source-network-drain system.
It is apparent that the width of the resonance peak is rather small,
reflecting the narrow transition bands in the source and drain channels.
Detuning in the transition bands between the source and the drain rapidly reduces both the peak height and width.
In the resonant case, $\epsilon_{s}=\epsilon_{d}$, the transmission spectrum is almost unaffected by
the size of the energetic fluctuations in the drain, as can be seen by comparing the
central peaks in Figs.~2(a) and 2(b). On the contrary, in the detuning case, the energetic disorder
helps enhance energy transfers by increasing both the height and width of the resonance peak. It is interesting to note that even
though the transition bands in the source and drain channels are in complete
detuning, the transmission peak is only partially suppressed, but
does not vanish. This means that such a pigment
network structure with multiple antennas and reaction centers is a
highly robust system for photosynthesis.

\begin{figure}
\includegraphics[width=13.5cm]{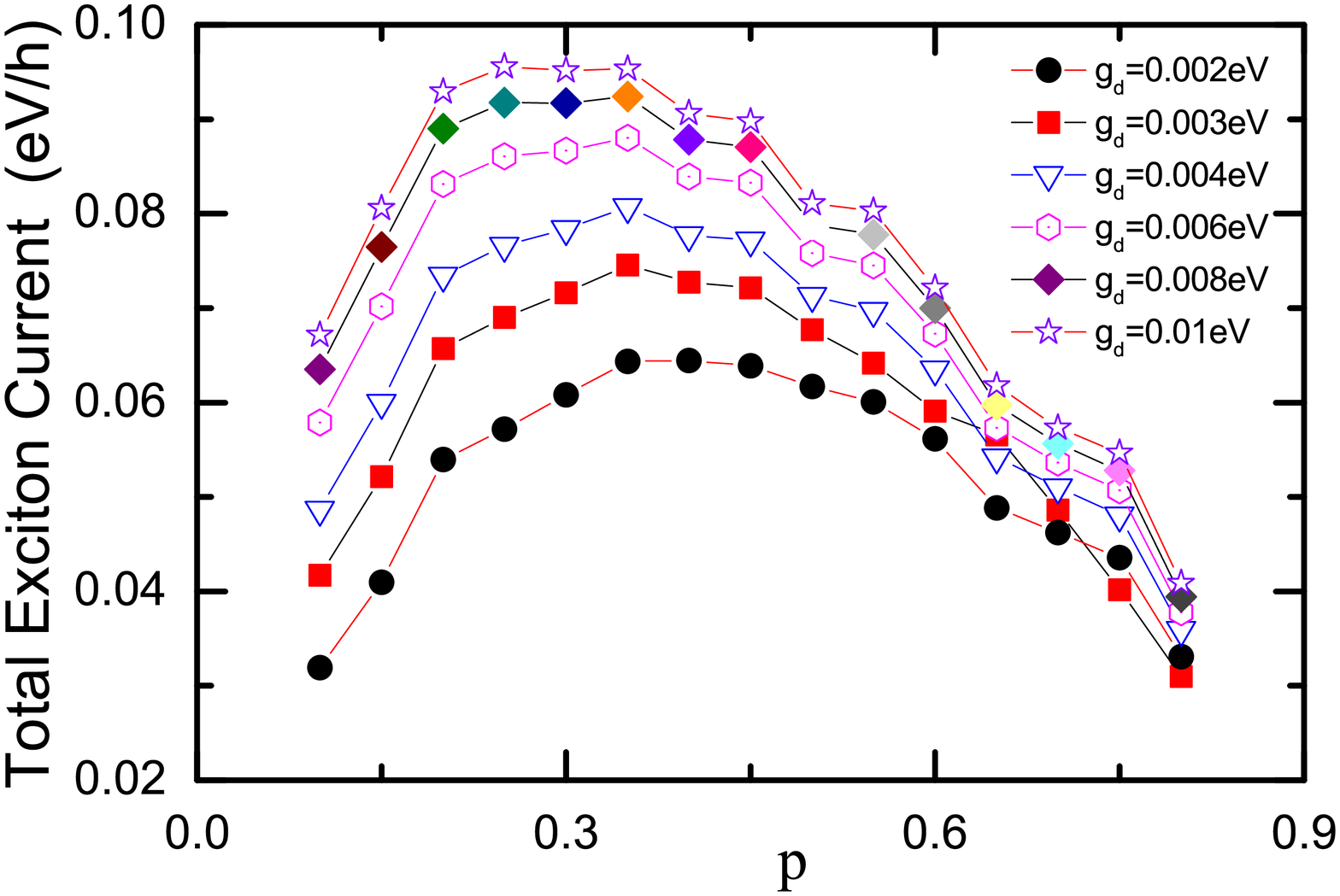}
\caption[FIG]{\label{FIG.3} Total exciton current as a function of
drain density $p$ for different values of $g_{d}$.
$\epsilon_{s}=\epsilon_{d}$, $w_d=0.002$eV, $w_s =0.001$eV, $g_{s}=0.001$eV, $J_0=0.01$eV, and $W=0.005$eV. For all combinations of $g_s$ and $g_d$ the curve of exciton current exhibits a single maximum at $p_c$, pointing to an optimal $p$ for energy transfer efficiency.}
\end{figure}

Now we investigate how the total exciton current, an indicator of the global energy transfer efficiency of the system, is affected when
the structure and control parameters of the pigment network are changed. In Fig.~3,
we show the exciton current as a function of the reaction center
density $p$ for various values of the trapping rate $g_{d}$ which controls the width of transition bands in reaction centers.
For a given set of $g_s$ and $g_d$, there exists
a critical reaction-center
density $p_{c}$ that maximizes the exciton current, i.e.,
at $p_{c}$, the excitons created in antennas can be most
efficiently transported to the reaction centers. Upon increasing the transition
band width $g_{d}$, $p_{c}$ is decreased but the exciton current is
increased, because for a larger reaction rate, a given number of excitons
can be collected by a smaller number of reaction centers. With the given form of $\eta_{ij}(E)$ in Eq.~(\ref{eta}), the exciton creation and trapping  rates in the antennas and the reaction centers may be controlled by $g_s$ and $g_d$, respectively. Thus results obtained here could be used to explain the fact that some purple bacteria develop more expansive antenna systems when cultured under low light conditions, as a single maximum in the exciton current for a given set of $g_s$ and $g_d$ indicates that the concentration $p$ may be adapted to the environment for optimal exciton transfer efficiency \cite{ada}. Our finding is also in qualitative agreement with the work of Fassioli {\it et al.} using a master equation approach, in which the efficiency can be optimized by changing the number ratio of LH2 and LH1 complexes (cf.~Fig.~4 of Ref.~\cite{lh1}). Nevertheless, the maxima
in Fig.~3 are not sharp peaks, pointing again to
the wide range of structural suitability for the network
to conduct photosynthetic energy transport.

To investigate how the energy transfer efficiency is affected by configurational disorder in our model, we plot in Fig.~4 the dependence of the total exciton current on the amplitudes of disorder in the pigment network, the sources, and the drains.
It is interesting to note that the exciton transfer  efficiency can be substantially enhanced
by increasing the energetic disorder in the antennas and the reaction centers.
The major efficiency bottlenecks are the low creation and annihilation rates in the antennas and reaction centers
that lead to a very narrow transmission resonance centered at the exciton energies as shown in Fig.~2.
Although the restriction of the narrow band on the efficiency can be partially lifted by increasing the number of channels, there is still a limitation on the maximum current through the drain. By introducing the disorder in the exciton energy,
the resonant energies spread over a wider range, and consequently,
the transmission resonance is widened and the exciton current increases.
In the work of Plenio and co-workers \cite{plenio}, the efficiency can be enhanced by increasing the environment noise in some parameter regimes which correspond to $J_{12}$ of about 100 cm$^{-1}$ and a full width at half maximum of the spectral density of the noise from 10 cm$^{-1}$ to 50 cm$^{-1}$
(cf.~Fig.~4 of Ref.~\cite{plenio}). This is comparable to $J_0 \sim 0.01$ eV and $w_{s(d)}$ from 0.0012 eV to 0.0062 eV in Fig. 4.
Computer simulations carried out by Mohseni {\it et al.} also suggest that random noise in the environment might actually enhance the efficiency of the energy transfer in photosynthesis rather than degrade it \cite{the15}.
On the other hand, the transfer rate in the network is usually much larger than
the creation and annihilation rates in the antennas and in the reaction centers, i.e.,
$J_0 \gg g_s$ and $J_0 \gg g_d$, so the excitonic quantum coherence
can be easily preserved during the exciton transfer across the network, and the transfer efficiency has a
very weak dependence on the network disorder $W$ (i.e., the variance in the NN coupling in the network),
as shown by the flat $I(W)$ curve in Fig.~4. This helps explain why a noisy environment can still
preserve quantum coherence and the randomness in antennas and reaction centers even enhances the energy transfer efficiency.

\begin{figure}
\includegraphics[width=13.5cm]{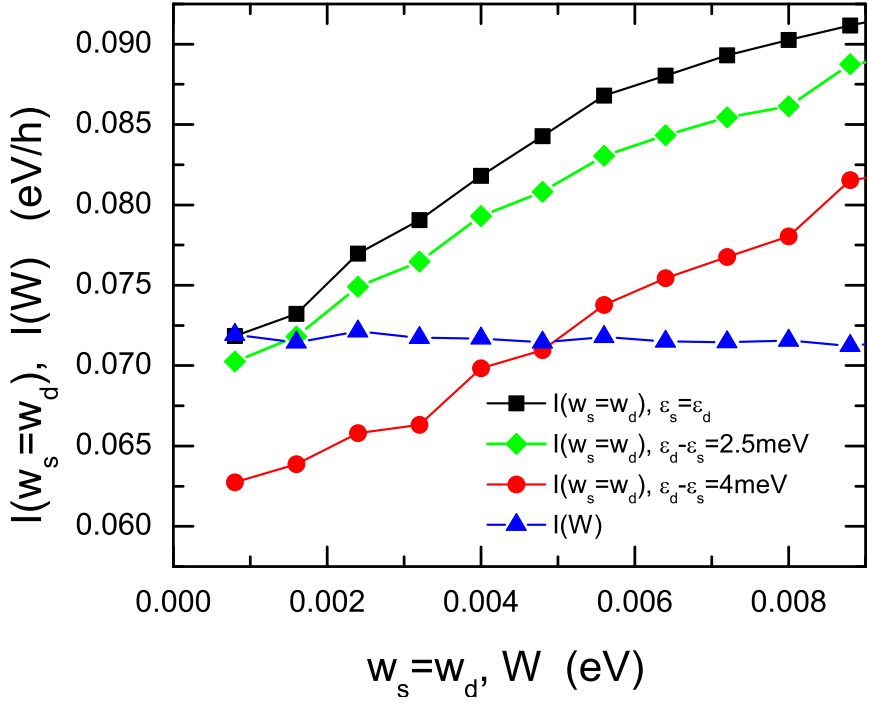}
\caption[FIG]{\label{FIG.4} Dependence of the total exciton current as a function of the degrees of disorder $w_d$ ($=w_s$) and $W$. For $I(w_s=w_d)$, $W=5$meV. For $I(W)$, $w_d=2$meV, $w_s=1$meV, and $\epsilon_s= \epsilon_d$. Other parameters are: $p=0.3$, $\epsilon_s= 1.5$eV,
$g_{d}=0.003$eV, $g_{s}=0.001$eV, and $J_0=0.01$eV. The exciton current is enhanced by energetic fluctuations in the antennas and the reaction centers, but it is almost unaffected by disorder in the NN hopping integral of the pigment network.}
\end{figure}

\section{Effect of thermally excited phonons}

Previously, the effect of phonons on the energy transfer process was excluded in the discussion. However, in realistic photosynthetic systems, thermally excited phonons are always present and are believed to play important roles. In this section we investigate the effects of phonons and thermal excitations on the energy transport efficiency. The phonon-related Hamiltonian can be written as \cite{ph1,ph2,EET}
\begin{equation}
\label{pss}
   H_{\rm ph} = \sum_{k} \hbar \omega_k  b_k^\dag b_k + \sum_k \sum_{i} \lambda_{k,i} a^\dag_{i,0}a_{i,0} ( b^\dag_k +b_k),
   \end{equation}
where $b_k$ ($b_k^\dag$) is the phonon annihilation (creation) operator for phonon mode $k$, $\omega_k$ is the corresponding phonon frequency, and $\lambda_{k,i}$ labels the exciton-phonon interaction strength at site $i$ for phonon mode $k$. A wave function in the phonon subspace can be written as
\begin{equation}
    |\varphi_m \rangle = \sum_{\{n_k\}} c_{m, \{ n_k \}} |\phi_{\{n_k\}} \rangle,
\end{equation}
where
\[
   |\phi_{\{n_k\}} \rangle = \prod_k \frac{ (b^\dag_k)^{n_k}}{\sqrt{n_k !}} |0 \rangle
   \]
with $\{ n_k\}$ being a set of phonon numbers in all modes, and $c_{m, \{ n_k \}}$ is the coefficient in the superposition.
In the coupled exciton-phonon system, the wave function in Eq. (\ref{wave}) can be extended to include the phonon Hilbert subspace
\begin{equation}
\label{wavep}
  \Psi_{i,\{ n_k \}} (E) =  \Psi_{i, \{n_k\}}^{\rm in}(E) + \sum_{j \in {\cal D}} \sum_{ \{ n_k^{'} \}}  \Psi_{i, \{n_k \}; j, \{n_k^{'} \}}^{\rm out}(E)
  +\sum_{j \in {\cal S}} \sum_{ \{ n_k^{'} \}} \Psi_{i, \{n_k \}; j, \{n_k^{'} \}}^{\rm ref} (E),
  \end{equation}
where
\begin{equation}
\label{ps1}
  \Psi_{i,\{ n_k\}}^{\rm in}(E) = \sum_{n=0}^\infty \exp[-i k_{i, \{ n_k\}}(E) n] a^\dag_{i,n} | 0\rangle \otimes |\phi_{\{n_k\}} \rangle ,
  \end{equation}
and
\begin{equation}
\label{ps2}
  \Psi_{i, \{ n_k \} ; j, \{ n_k^{'} \} }^{\rm out(ref)}(E) = \sum_{n=0}^\infty t(r)_{i , \{ n_k \}; j, \{ n_k^{'} \} }(E) \exp[i k_{ j, \{ n_k^{'} \}} (E) n] a^\dag_{j,n} | 0\rangle \otimes |\phi_{\{n_k^{'}\}} \rangle \,\,\,{\rm   for   }\,\,\, j \in {\cal D}({\cal S}).
  \end{equation}
Here $\Psi_{i,\{ n_k \}} (E)$ is the complete wave function for an exciton entering from channel $i$ clothed by a phonon cloud $|\phi_{\{n_k\}} \rangle$, and $E$ now labels the total energy of the exciton and companying phonons. The first, second, and third terms in Eq.~(\ref{wavep}) are the incident, transmitted and reflected parts of the wave function in different channels and with different phonon states. Given the total energy and the phonon energies the wave vector is calculated as
$k_{i,\{ n_k\}}  (E) = \arccos [(E-\epsilon_i- \sum_k n_k \hbar \omega_k )/{2 g_i}] $.
For an exciton-phonon composite with total energy, $E$
 $t_{i , \{ n_k \}; j, \{ n_k^{'} \} }(E)$ and $r_{i , \{ n_k \}; j, \{ n_k^{'} \} }(E)$ are transmission and reflection amplitudes from channel $i$ with phonon state $ |\phi_{\{n_k\}} \rangle$ to channel $j$ with phonon state $ |\phi_{\{n_k^{'}\}} \rangle$, respectively, which can be determined from the Schr\"{o}dinger equation
\begin{equation}
     (H_0 +H_{ph}) \Psi_{i,\{ n_k \}} (E) = E  \Psi_{i,\{ n_k \}} (E)
\end{equation}
with a much-enlarged Hilbert space.
By substituting Eqs. (\ref{ps1}) and (\ref{ps2}) into Eq. (\ref{pss}), one may obtain terms such as
\[
  c_{j,1} \left( \sum_{n=0}^\infty \exp[i k_{j,1} n] a^\dag_{j,n} \right)
  \otimes b_1^\dag |0 \rangle +c_{j,1} \left( \sum_{n=0}^\infty \exp[i k_{j,2} n] a^\dag_{j,n} \right)
  \otimes b_2^\dag |0 \rangle
  \]
in output channel $j$, where $k_{j,1(2)} =\arccos [(E-\epsilon_j - \hbar \omega_{1(2)})/2g_j]$, and $c_{j,1(2)}$ is the coefficient related to the corresponding transmission amplitudes. Obviously these terms cannot be factorized if $k_{j,1} \neq k_{j,2}$. Thus, in general, the exciton and phonon states are entangled in an output channel even if the two are not in the incoming states.

As the creation and annihilation of excitons in the antennas and the reaction centers is very slow, it is reasonable to assume that when an exciton is created in an antenna, the phonon equilibrium distribution in the network has recovered from the perturbation by the movement of the previous exciton. Thus the probability of the phonon state being $|\phi_{\{ n_k\}}\rangle$ which accompanies an incident exciton is
\begin{equation}
   P_{\{ n_k \}}(T) = \frac{1}{Z} \exp[ - \sum_k n_k \hbar \omega_k /k_BT],
   \end{equation}
where the partition function is defined as
\[
   Z= \sum_{\{ n_k\}} \exp[ - \sum_k n_k \hbar \omega_k /k_BT].
   \]
Therefore, by including the interactions with phonons, the total exciton current at temperature $T$ can be calculated as
\begin{equation}
     I(T)= \frac{1}{h} \sum_{ i\in {\cal S}, j \in
    {\cal D}}\sum_{ \{ n_k \}, \{ n_k^{'} \}} \int_{{\cal E}_{i, \{n_k\}; \, j, \{ n_k^{'} \} }} dE \frac{ P_{\{ n_k \}}(T) |t_{i, \{n_k \}; \, j, \{ n_k^{'} \}}(E) |^2 |g_j \sin
    k_{j,\{ n_k^{'} \}} (E)|}{|g_i \sin k_{i, \{ n_k \}} (E)|},
    \end{equation}
where the integration range ${\cal E}_{i, \{ n_k \}; \, j, \{ n_k^{'} \} }$ is the span of the total energy $E$ within which both
$k_{i, \{ n_k \}}(E)$ and $k_{j, \{n_k^{'} \}} (E)$ are real.

\begin{figure}
\includegraphics[width=13.5cm]{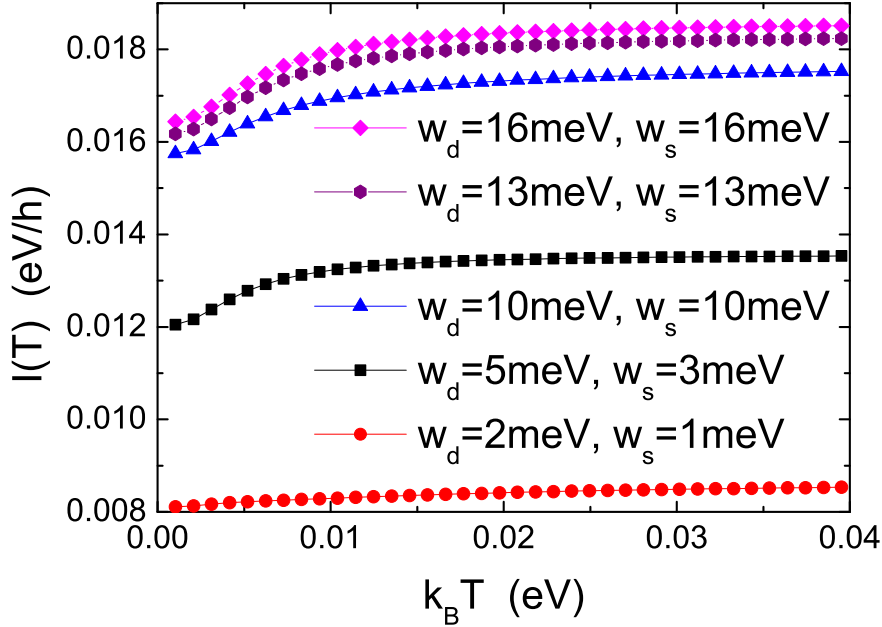}
\caption[FIG]{\label{FIG.5} Temperature dependence of the total exciton current for various amplitudes of energetic disorder in the sources and the drains. Other parameters are: $p=0.3$, $\epsilon_s= 1.5$eV, $\epsilon_d -\epsilon_s =2$meV,
$g_{d}=1$meV, $g_{s}=0.5$meV, $J_0=0.01$eV, $W=5$meV, $\omega_c =60$meV, and $\lambda_0 =1$meV. For a given value of exciton-phonon coupling strength, e.g., $\lambda_0 =1$meV, the exciton current increases with the increasing temperature for all amplitudes of energetic disorder.}
\end{figure}

Because of computational constraints, only a restricted number of phonon states are included. In the calculation, we adopt a uniform phonon frequency distribution in interval $[0,\omega_c]$, which is similar to a simplified form of spectral density in Ref.~\cite{ww1}. We adopt ten equally spaced modes in this interval, and the phonon states with energies much greater than $k_BT$ are not included. The interaction strengths $\lambda_{k,i}$ are assumed to be independent of the modes, i.e., $\lambda_{k,i} = \lambda_0$.
Calculated temperature dependence of the exciton current is plotted in Fig.~5 for various amplitudes of energetic disorder in the sources and the drains, and in Fig.~6 for various strengths of exciton-phonon coupling. It is interesting to note that, similar to the aforementioned disorder effect, in most cases by increasing the temperature the exciton current is enhanced, and eventually reaches saturation at room temperature.
This means that the thermal excitations may actually favor energy transport in photosynthetic systems. The opposite trend, that the current decreases with increasing temperature, only occurs in a very narrow range of coupling strength, i.e., 6 meV $ \leq \lambda_0 \leq $ 7 meV.
This result is in agreement with
that obtained by Mohseni {\it et al.} \cite{the15} who has shown that an effective interplay
between free Hamiltonian evolution and thermal fluctuations in the environment leads to a
substantial increase in energy transfer efficiency (from about 70\% to 99\%).
Using a phonon mode of 180 cm$^{-1}$ ($\sim 22$meV) and an exciton-phonon coupling strength of $ \sim 10$ meV, Plenio and coworkers showed that the coupling to phonons may significantly increase energy transfer efficiency (cf.~Fig.~7 of Ref. \cite{Chin}). Similar values of the exciton-phonon coupling strength can also be found in Fig.~6 where  an enhancement of the exciton current with increasing temperature is shown.
The phonon modes which interact with excitons can play the role of scatterers for the exciton transport, impeding the energy transmission, but they can also provide alternative paths for exciton transfers, therefore promoting energy transport. Our results show that at low temperatures the latter effect is dominant in most parameter regimes, but at high temperatures the two effects are in balance, yielding a temperature-independent exciton current.

\begin{figure}
\includegraphics[width=13.5cm]{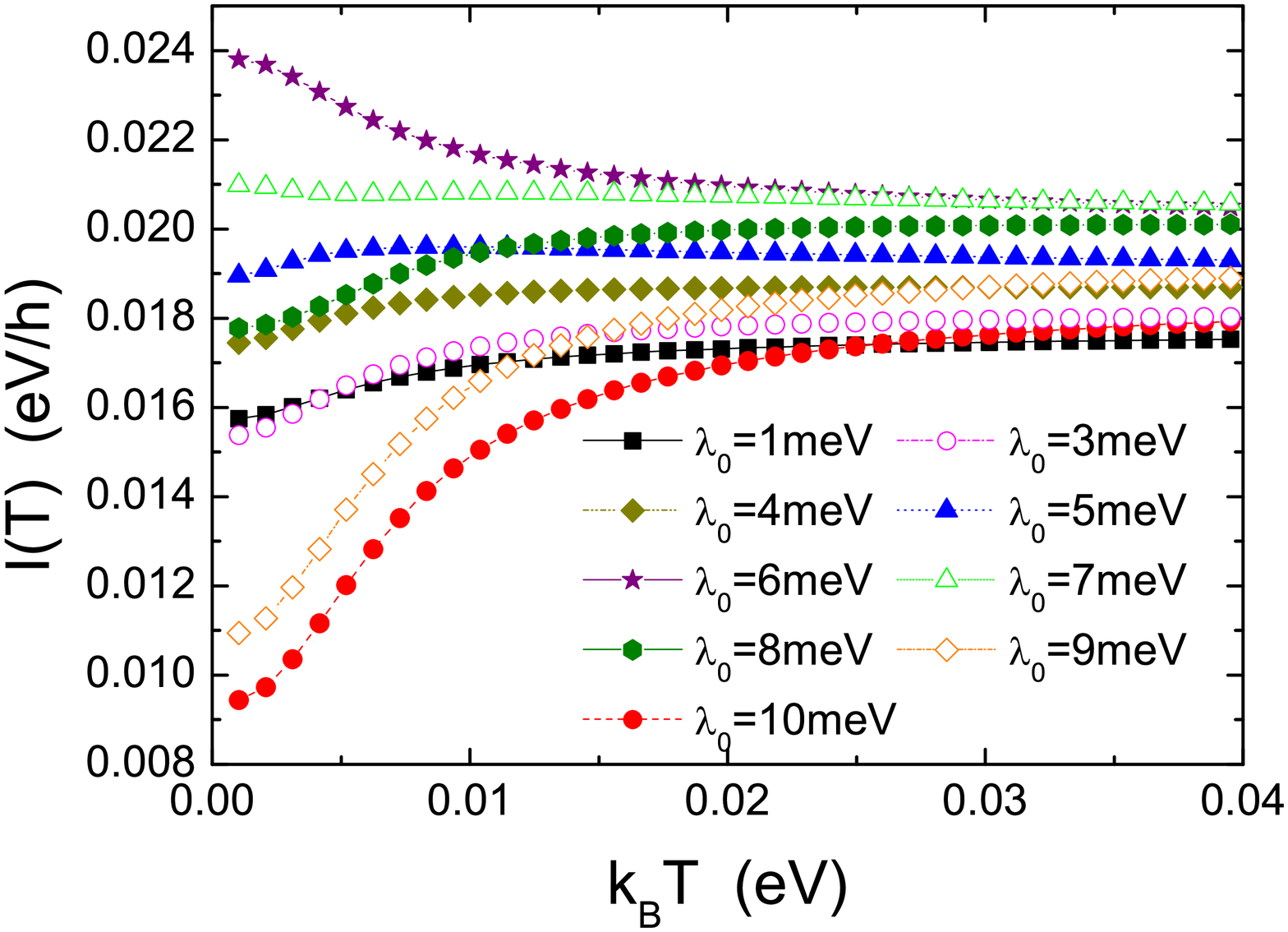}
\caption[FIG]{\label{FIG.6} Temperature dependence of the total exciton current for various strengths of exciton-phonon coupling. Other parameters are: $p=0.3$, $\epsilon_s= 1.5$eV, $\epsilon_d -\epsilon_s =2$meV, $w_d = w_s=10$meV,
$g_{d}=1$meV, $g_{s}=0.5$meV, $J_0=0.01$eV, $W=5$meV, and $\omega_c =60$meV. The exciton current is found to increase with the temperature for all exciton-phonon coupling strength except the narrow bracket from 6 meV to 7 meV.}
\end{figure}

\section{Conclusions}

 In this work a source-network-drain exciton Hamiltonian is introduced to investigate the energy transport processes in photosynthetic systems. The excitons are electronic energy carriers, and the creation and annihilation of the excitons in antennas and reaction centers are described by source and drain channels, respectively. The creation and annihilation rates in antennas and reaction centers are characterized by the band widths in corresponding channels and by the statistical factors of excitons depending on light intensity, adsorption cross section in antennas and the reopening timescale in reaction centers. The static disorder in antennas and reaction centers is taken into account by introducing randomness of corresponding parameters in the Hamiltonian, while the effects of thermally excited phonons are described by exciton-phonon interactions. As the total  Hamiltonian is Hermitian and the Schr\"{o}dinger equation is solved for the combined wave function of excitons and phonons, our treatments here are fully quantum mechanical and the effect of coherence is included accordingly. The exciton current, an efficiency indicator of the energy transport, can be calculated using the Landauer-Buttiker formula. Our results show that the exciton current may be enhanced by increasing the static disorder and increasing the temperature in many parameter regimes, suggesting that the coherent energy transport in light-harvesting systems is robust against most environmental noises, a conclusion that is in agreement with previous findings in the literature. Furthermore, the current approach provides a novel, flexible platform to investigate complicated interplays among various configurational, quantum and thermal factors in the energy transport processes of natural and artificial photosynthetic systems.

 Previously, many theoretical efforts were based upon solving the phenomenological Lindblad master equations with built-in dissipation. In this work we are interested in the steady state in which excitons flow continuously to the reaction centers, and our results are not affected by initial conditions and transient behavior. It is also possible to find the steady-state density matrix for a single-particle Lindblad equation. We note that the exciton life time is in the order of nanoseconds while it takes hundreds of femtoseconds for an exciton to be transferred to a reaction center. This huge difference in timescale indicates that exciton decay can be neglected when considering excitonic energy transfers from antennas to reaction centers, therefore lending support to the Landauer-Buttiker scheme adopted here.
Our approach assumes the single-exciton picture in which the Schr\"{o}dinger equation is solved, and  the applicability of this basic assumption to photosynthetic systems is justified by slow injection and fast transport of excitons in the pigment network. In a master-equation approach, energy transfer efficiency is measured by  quantities such as the exciton trapping probability, while in the Landauer-Buttiker scheme it is described by the exciton current.
Furthermore, the total exciton current also provides a global description of the photosynthetic system with multi-channel inputs and outputs and interferences among them.

\section*{Acknowledgments}

Support from the Singapore National Research Foundation
through the Competitive Research Programme
(CRP) under Project No.~NRF-CRP5-2009-04 is
gratefully acknowledged.
This work is also supported in part by the State Key Programs for Basic Research
of China (Grant No. 2011CB922102), and by National
Foundation of Natural Science in China Grant No.
61076094.

\end{document}